\documentclass[11pt,a4paper]{article}
\usepackage{jheppub}
\newlength{\figurewidth}
\newcommand{\beq}{\begin{equation}}
\newcommand{\eeq}{\end{equation}}
\newcommand{\bea}{\begin{eqnarray}}
\newcommand{\eea}{\end{eqnarray}}
\newcommand{\ba}{\begin{array}}
\newcommand{\ea}{\end{array}}

\newcommand{\mn}{{\mu\nu}}
\newcommand{\pt}{\partial}
\newcommand{\pd}{(2\pi)^d}
\newcommand{\al}{\alpha}
\newcommand{\bt}{\beta}

\newcommand{\ep}{\epsilon}

\newcommand{\G}{\Gamma}

\newcommand{\de}{\delta}
\newcommand{\D}{\Delta}
\newcommand{\sg}{\sigma}

\title{Charge Renormalization due to Graviton Loops}
\author{Gaurav Narain}
\author{Ramesh Anishetty}
\affiliation{The Institute of Mathematical Sciences,
Taramani, Chennai 600113, India.}
\emailAdd{gaunarain@imsc.res.in}
\emailAdd{ramesha@imsc.res.in}
\abstract{
The leading term in the gauge coupling beta function 
comes due to interaction of gauge field with gravitons. 
It is shown to be a universal quantity for all gauge theories. 
At one-loop it is found to be zero in four dimensions. This is independent of 
the gravity action with metric as the field variable, gauge fixing condition
and regularization scheme. This term being universally same for all gauge 
groups is further studied in the case of abelian gauge theories, where due to self-duality
this term is shown to be zero to all loops, on-shell. Consequences of this are discussed.
}
\begin{document}
\maketitle
\section{Introduction.}
\label{intro}

Asymptotic freedom at high energies is a property enjoyed 
by non-abelian gauge theories due to their unique charged self interaction. 
Sufficiently small number of interactions due to charged boson or fermions 
do not alter this property. Gauge fields interact with the metric
fluctuation due to their energy, which is weak at energies below Planck scale.
However at arbitrarily high energies quantum effects due to gravity cannot be 
ignored, which might affect asymptotic freedom of non-abelian or the Landau singularity 
of abelian gauge theories. Keeping this in mind it is important to know how the 
high energy behavior of gauge couplings is affected when quantum 
gravity effects are taken in to account?

Quantum gravity effects on charge renormalization were first 
discussed in \cite{Deser1, Deser2, Deser3, Deser4} within perturbation theory using
dimensional regularization and concluded that at one-loop there 
is no modifications to running gauge coupling due to gravitons. 
This result was often suspected 
to be a consequence of massless nature of gluons/photons and 
gravitons, and the way dimensional regularization regulates 
quadratic divergences. The problem was re-examined 
in momentum cutoff regularization with $R_\xi$ type of 
gauge condition in \cite{Robinson}, and came to the conclusion 
that renormalized charge of gauge theories gets a nonzero correction 
from graviton loops and vanishes as power 
law (as opposed to logarithmic) well before Planck energies. Subsequently 
many authors have reanalyzed the issue using various ways in different 
gauge fixing choices and refuted the result of \cite{Robinson}. 
The investigations which find that there is no charge renormalization due 
to graviton loops: using harmonic type gauge condition and momentum cutoff
\cite{Pietrykowski}, dimensional regularization using 
gauge independent formulation of effective action \cite{Toms:2007},
using standard Feynman technique with both momentum cutoff 
and dimensional regularization \cite{Rodigast2008}, 
using Functional renormalization group \cite{Folkerts}. The
literature which finds a nonzero graviton contribution to the 
running of gauge couplings: loop regularization \cite{TangWu1},
in presence of cosmological constant
using Vilkowisky-DeWitt technique \cite{TomsCosmo1, TomsCosmo2},
studying quadratic divergences using Vilkowisky-DeWitt technique
\cite{TomsQuad1, TomsQuad2}, using Functional renormalization group equation \cite{Daum}.
In all these studies, quantum field theory of Einstein-Hilbert (EH) gravity was
used to obtain charge renormalization due to graviton loops.
In \cite{Fradkin}\footnote{Here the authors commented about it in the footnote calling
it a `mysterious cancellation'} the authors studied the coupling of fourth 
order higher derivative gravity with gauge field 
(which is renormalizable to all loops) showing that at one-loop
no charge renormalization due to graviton loops happens. 
In \cite{Robinson, Pietrykowski, Toms:2007, Rodigast2008, TangWu1, 
TomsCosmo1, TomsCosmo2, TomsQuad1, TomsQuad2, Fradkin} the computation was performed 
in the framework of Effective field theory \cite{Donoghue1, Donoghue2, Donoghue3}, where it is believed that 
any sensible theory of quantum gravity must have EH
gravity as its low energy limit. Therefore it is natural to inquire in those settings,
what is the one-loop graviton correction to the running of gauge coupling? 
In \cite{Daum, Folkerts} however Functional renormalization 
group was used to study the problem in the spirit of asymptotic safety scenario 
\cite{Weinberg, AS_rev1, AS_rev2, AS_rev3, AS_rev4}.
 
Higher derivative gravity in four dimensions is perturbatively renormalizable 
to all loops \cite{Stelle}, which has recently been shown to be unitary \cite{NarainA1, NarainA2}.
The higher-derivative gravity action considered is,
\beq
\label{eq:hdgact}
S
= \int \frac{{\rm d}^4x \sqrt{-g}}{16 \pi G} \left[
 -R + \frac{\omega R^2}{6 M^2}
- \frac{R_\mn R^\mn - 
\frac{1}{3}R^2}{M^2}
\right] \, ,
\eeq
where $G$ is the gravitation Newton's constant, $M$ is a mass 
parameter and $\omega$ is a positive real dimensionless constant. 
Under quantum corrections the coupling parameters runs with energies.
It has been found that the running coupling $G$ remains small for all 
energies and hence within the perturbative domain. 
Furthermore it vanishes at some finite energy albeit larger than the 
Planck energy. The coupled system of matter fields with higher-derivative 
gravity being perturbatively renormalizable to all loops \cite{Fradkin}, 
hardly effects the above behavior of the running of $G$. Therefore in this 
setting it is natural to wonder how charge renormalization of gauge theories 
will get effected once quantum corrections from gravity are taken into account. 
It should be noted that this system being perturbatively renormalizable to all
loops and being free from quadratic divergences, evades the criticism 
raised in \cite{Donoghue4}, which holds in the case of effective field 
theories of gravity like Einstein-Hilbert gravity where the presence of 
quadratic divergences  does not allow one to give proper meaning to the 
quantum corrections to the coupling. 

Generally when gauge field is coupled with gravity, then it is found that
the beta function for the gauge field coupling 
in perturbation theory is governed by a term 
called `$a$' to all loops, which has the dominant contribution for 
small gauge coupling constant and is universally the same for all 
gauge groups including the abelian gauge theories. However `$a$' can depend 
on the parameters of the pure gravity sector. Potentially `a' term can change 
the asymptotic freedom result as alleged in \cite{Robinson, TomsQuad1, TomsQuad2}. In the 
following we define `$a$' terms more precisely and calculates it to one-loop.
Using duality transformations and the fact that the two theories are 
equivalent under quantum corrections on-shell \cite{Duality1, Duality2, FradkinD}, 
we conclude that this `$a$'-term vanishes to all loops on-shell.
 
The non-abelian gauge field action is given by, 
\beq
\label{eq:gaugeact}
S_{\rm gauge}=
- \frac{1}{4e^2} \int {\rm d}^4x \sqrt{-g} 
g^{\mu\al} g^{\nu\bt} F_{\mn}^a F_{\al\bt}^{a} \, ,
\eeq
where $e$ is the $SU(N)$-gauge coupling, $F_{\mu\nu}^a
=\partial_{\mu} A_{\nu}^a - \partial_{\nu}A_{\mu}^a
+ f^{abc}A_{\mu}^bA_{\nu}^c$, and $A_{\mu}^a$ is the 
gauge vector potential. We define the quantum theory 
of gravity and gauge fields by Feynman path-integral 
corresponding to the action given in eq. (\ref{eq:hdgact} 
and \ref{eq:gaugeact}) which is perturbatively 
renormalizable to all loops \cite{Fradkin, Moriya}.
We use background field method and do the 
appropriate gauge fixing for gravity and gauge sector.
In our case in particular, the background gauge and gravity fixings guarantee 
gauge invariant effective action by using $4-\ep$ dimensional 
regularization scheme \cite{DeWitt1, DeWitt2, tHooft, Abbott}.
The running of gauge coupling constant satisfies the following 
generic equation,
\beq
\label{eq:beta_gauge_gr}
\frac{{\rm d}}{{\rm d} t} \left(
\frac{1}{e^2} \right) = \frac{a(M^2G, \omega)}{e^2} 
+ b(e^2, M^2G, \omega) \, ,
\eeq
where $t= \ln (\mu/\mu_0)$, the function `$a$' is independent of 
$e^2$ and the gauge group but can depend upon couplings 
in gravity sector, while `$b$' depends on gauge group and 
all couplings present in the theory. The above definition is particularly useful 
when $e^2$ is small, because by construction `$b$' is a regular function 
of $e^2$ at $e^2=0$. Feynman perturbation theory naturally gives the above. 

At one-loop the two diagrams that are giving the gravitational contribution to the running 
of gauge coupling are shown in Fig. \ref{fig1}. 
Concentrating on $e^2$ dependence, we notice that any vertex which 
involves gluon line gives $1/e^2$ while every gauge propagator 
in the loop gives $e^2$, hence the one-loop diagrams are proportional 
to $1/e^2$. Consequently all the one-loop diagrams contribute to 
`$a$' term alone in eq. (\ref{eq:beta_gauge_gr}). 

Now we would like to address `$a$' term to all loop orders 
in the expansion. First we note that any diagram with triple 
or quadrupole gauge vertices, by simple power counting of $e^2$ automatically 
contributes only to `$b$' term. So effectively we can ignore them 
to evaluate `$a$' term. By doing so it is evident that all such diagrams 
are also common to $U(1)$ gauge theory. Ghosts and presence of matter 
fields in the theory again contributes only to `$b$' term. Therefore 
`$a$' term is universal to all gauge theories and is the most 
dominant term for small gauge coupling constant. The universality 
suggests that it is a manifestation of the fact that the metric fluctuations
interact universally to all gauge fields via its energy. 

The formal solution of eq. (\ref{eq:beta_gauge_gr}) can be written as,
\beq
\label{eg:e^2run}
\frac{1}{e^2} = e^{ \int_0^t a \, {\rm d} t^{\prime}} \biggl(
\frac{1}{e_0^2} + \int_0^t {\rm d}t^{\prime} 
\, b \, e^{- \int_0^{t^{\prime}} a \, {\rm d} t^{\prime\prime}}
\biggr)
\eeq
It is evident that for small $e^2$ the running of $e^2$ depends 
more dramatically on the sign of $a$. If $a$ is negative then 
$e^2$ diverges as $e^{|a| t}$ for large $t$. 
If $a$ is positive, then for large $t$, $e^2$ vanishes 
faster than $e^{-|a| t}$ as considered in \cite{Robinson, TomsQuad1, TomsQuad2}.
If $a=0$, then the standard behavior of the
running of gauge coupling qualitatively holds. The above equation is valid to any order in the loop 
expansion. So the qualitative behavior of $e^2$ namely asymptotic freedom 
can remain unaltered if $a=0$ to all loops. We examine this in the following. 

The outline of paper is: in section. \ref{oneloop}, we compute the one-loop quantum gravity 
contribution to the charge renormalization in higher derivative gravity, in section. \ref{duality} 
we use duality transformations to study the `$a$' term of the gauge beta function 
to all loops in abelian gauge theories, in section. \ref{arbitGR} we compute the one-loop quantum 
gravity contribution to charge renormalization in arbitrary metric theory of gravity, we conclude 
with a discussion in section. \ref{discuss}.

\section{One-loop Computation.}
\label{oneloop}

\begin{figure}
\centerline{
\begin{minipage}[t]{3.0in}
\vspace{0pt}
\centering
\includegraphics[width=3.0in]{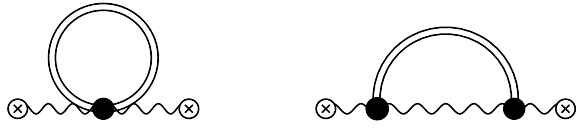}
\end{minipage}
}
 \caption[]{
One loop gravitational correction to gauge coupling.
Black dots represent the bare vertices, while wavy 
lines attached to circle with a cross represent external legs.
  }
\label{fig1}
\vspace{-5mm}
\end{figure}

We evaluate explicitly the diagrams of Fig. \ref{fig1}. Following 
\cite{NarainA1, NarainA2} we choose flat background 
and a physical gauge choice for graviton
namely the Landau gauge ($\pt_\mu h^\mn=0$) 
wherein the graviton propagator is given by,
\beq
\label{eq:grav_prop1}
D_{\mu\nu,\rho\sg} =
\frac{i \, 16 \pi}{(2 \pi)^4} (M^2G) 
\Biggl[
\frac{1}{\omega} \frac{(P_s)_{\mn,\al\bt}}{q^2(q^2 - \frac{M^2}{\omega})}
- \frac{2 \, (P_2)_{\mn,\al\bt}}{q^2(q^2 - M^2)}
\Biggr] 
\eeq
where $q$ is the momentum of fluctuating metric field. The various 
spin projectors are defined in terms of the following two
projectors in arbitrary dimensions: $L_{\mn}=q_{\mu}q_{\nu}/q^2$ and
$T_{\mn}=\eta_{\mn} - L_{\mn}$;
\bea
\label{eq:proj} 
&&
(P_2)_{\mn,\al\bt} = 
\frac{1}{2} \left[ T_{\mu\al} T_{\nu\bt} + 
T_{\mu\bt}T_{\nu\al} \right] 
- \frac{1}{d-1} T_{\mn}T_{\al\bt} \, ,
\notag \\
&&
(P_s)_{\mn,\al\bt} = 
\frac{1}{d-1} T_\mn \, T_{\al\bt} \, .
\eea
By simple partial fractionation it can be shown that this
corresponds to massless spin-2 gravitons, a massive 
(mass $M/\sqrt{\omega}$) spin-0 scalar and a massive 
(mass $M$) spin-2 mode identified to be a ghost
\cite{NarainA1, NarainA2}. 
The gauge propagator with momentum $p_\mu$ 
is taken to be,
\beq
\label{eq:gaugeprop}
\D_{\mn}^{ab} = \frac{-i e^2 \de^{ab}}{ (2 \pi)^4 p^2} \biggl[
\eta_{\mn} - (\al -1) \frac{p_\mu p_\nu}{p^2} 
\biggr] \, ,
\eeq
where $\al$ is an arbitrary gauge fixing parameter.
%
%
The contribution of each diagram in Fig. \ref{fig1} in arbitrary dimensions is,
\bea
\label{eq:tadbub}
\G_{\rm Tad} =&& 
- \frac{2 \pi M^2 G}{e^2 \, (4 \pi)^{\frac{d}{2}}}  \G \left(2 -\frac{d}{2} \right)
\biggl[
\frac{(d+1)(d^2 - 9d +12)}{d(d-1)} M^{d-4}
\notag \\
&&
+ \frac{1}{\omega} \frac{(d-5)(d^2 - 7d +8)}{2d (d-1)(d-2)} 
\left(\frac{M^2}{\omega} \right)^{\frac{d}{2}-2}
\biggr] \int {\rm d}^d x \, {\rm tr} F^2 \, ,
\notag \\
\G_{\rm Bub} = &&
 \frac{2 \pi M^2 G}{e^2 \, (4 \pi)^{\frac{d}{2}}}  \G \left(2 -\frac{d}{2} \right)
\biggl[
-\frac{8(d+1)}{d(d-1)} M^{d-4} 
\notag \\
&&
+ \frac{1}{\omega} \frac{2(d-3)^2}{d(d-1)(d-2)} 
\left(\frac{M^2}{\omega} \right)^{\frac{d}{2}-2}
\biggr]\int {\rm d}^d x \, {\rm tr} F^2  \, .
\eea
The $1/\ep$ pole of these diagrams in the $4-\ep$ dimensional 
regularization scheme is,
\beq
\label{eq:tadbubd4}
\G^{\rm Div}_{\rm Tad} = -\G^{\rm Div}_{\rm Bub} =
\frac{1}{\ep} \frac{M^2 G}{48 \pi \, g^2} \left(40 - \frac{1}{\omega} \right) 
\int {\rm d}^4 x \, {\rm tr} F^2 \, .
\eeq
The divergent contribution of the two diagrams given in eq. (\ref{eq:tadbubd4})
therefore exactly cancel each other in four space-time 
dimensions. Due to this the `$a$' term vanishes in the minimal 
subtraction scheme at one-loop. 
This vanishing of `$a$' in the context of higher 
derivative gravity was also observed in \cite{Fradkin}.

There are many instances in the literature where the same problem 
studied within the framework of EH gravity action found $a=0$. 
However in these cases of effective field theory, no clear meaning 
should be associated to such terms in the beta function \cite{Donoghue4}.
But still a natural question arises whether this is an 
accidental cancellation or there is a deeper principle which is at work.
This issue is all the more firmly raised in a renormalizable, unitary 
quantum gravity theory such as the one given by
eq. (\ref{eq:hdgact}) \cite{NarainA1, NarainA2}. Indeed 
we already found that this universal `$a$'-term is exactly quantitatively 
present even in simple $U(1)$ gauge theory coupled to gravity 
without any other matter fields. As `$a$' term is universal for 
all gauge theories, one is motivated to study this particular term of the beta function 
in more detail by working in the context of abelian gauge theories, where the `$b$' term 
is identically zero in the absence of matter. This setting is particularly simple and helps us to 
gain more insight in to the cause of cancellation of divergence, leading 
to vanishing of `$a$' term in the beta function. Therefore in the next section we 
will study the `$a$' term in abelian gauge theory to all loops.

\section{Duality symmetry.}
\label{duality}

In this section we give a formal argument to show 
that $a=0$ because of self duality property 
of the $U(1)$ gauge theory in four dimensions \cite{Duality1, Duality2, FradkinD}.
In \cite{FradkinD} quantum equivalence of dual theories
was shown to hold on-shell, thus the following argument 
to show that $a=0$ will hold on-shell. 
In this proof gravity action plays no significant role. 
Consider the Feynman path integral
\beq
\label{eq:U1pathint}
Z = \int {\cal D} g_\mn \, {\cal D} A_\rho 
e^{i(S_{\rm GR} + S_{\rm EM})},
\eeq
where $S_{\rm EM} = \left(-1/4e^2\right) \int {\rm d}^4 x \, \sqrt{-g}
\, g^{\mu\al} g^{\nu\bt} F_\mn F_{\al\bt}$, and 
$F_\mn= \partial_\mu A_\nu - \partial_\nu A_\mu$ and $S_{\rm GR}$
is some arbitrary renormalizable and unitary gravity action. For this theory 
the running of gauge coupling is given by,
\beq
\label{eq:e2beta}
\frac{{\rm d}}{{\rm d} t} \left( \frac{1}{e^2} \right)
= \frac{a(\cdots)}{e^2} \, ,
\eeq
where the dots indicate that the function `$a$' can depend 
upon parameters of $S_{\rm GR}$. 

Formally we rewrite the path-integral for $U(1)$ gauge field
by making use of auxiliary tensor field $B_\mn$. 
\bea
\label{eq:dualtrans}
\int {\cal D} A_{\mu} e^{i S_{\rm EM}} = \int {\cal D}B_\mn \, 
{\cal D} A_{\mu} e^{i S_B} \, \left[
\det (e^2 G^{\mn,\al\bt})
\right]^{\frac{1}{2}} \, ,
\eea
where 
\bea
\label{dualtrans1}
&&
S_B =e^2 \int {\rm d}^4 x \sqrt{-g}  g^{\mu\al} g^{\nu\bt}
B_\mn B_{\al\bt} + \int {\rm d}^4 x \, \ep^{\mn\al\bt}
B_\mn \pt_{\al} A_\bt  \, , \\
&&
\label{eq:dualtrans2}
G^{\mn,\al\bt} = \sqrt{-g} (g^{\mu\al} g^{\nu\bt}- g^{\mu\bt} g^{\nu\al}) \, .
\eea
$G^{\mn,\al\bt}$ is anti-symmetric in $(\mn)$ and 
$(\al\bt)$, and $\ep^{\mn\al\bt}$ is a four dimensional tensor density 
of weight $-1$. On integrating the $A$ field on the rhs 
of eq. (\ref{eq:dualtrans}) we get a delta function 
$\de \left( \ep^{\mn\al\bt} \pt_{\al} B_\mn \right)$,
which constraints the $B$ field. On a classical level, this constraint 
arises from the equation of motion for $A_\mu$ field of the action given in 
eq. (\ref{dualtrans1}). The solution $B_\mn = \pt_\mu b_\nu
-\pt_\nu b_\mu$ satisfies the constraint. Then eq. (\ref{eq:dualtrans}) becomes,
\bea
\label{eq:dualpathint}
Z= C \int {\cal D} g_\mn \, {\cal D} b_{\mu}
e^{i (S_{\rm GR} + \bar{S}_{\rm EM})} 
\left[
\det (e^2 G^{\mn,\al\bt} )\right]^{\frac{1}{2}} 
\, ,
\eea
where $C$ is a constant and $b_\mu$ is the dual 
of $A_\mu$ whose action is given by,
\beq
\label{eq:dualact}
\bar{S}_{\rm EM} = e^2 \int {\rm d}^4 x \, \sqrt{-g} 
g^{\mu\al} g^{\nu\bt} (\pt_\mu b_\nu - \pt_\nu b_\mu) 
(\pt_\al b_\bt - \pt_\bt b_\al) \, .
\eeq
Here we have implemented the duality transformation
in the presence of metric \cite{Duality1, Duality2}. 
This dual action also has $U(1)$ gauge invariance.
Next we note that the matrix $e^2 G^{\mn,\al\bt}$
which is anti-symmetric in $(\mn)$ and $(\al\bt)$,
and is an ultra-local $d(d-1)/2 \times d(d-1)/2$ matrix which is general 
co-ordinate invariant. Hence the determinant can only be proportional 
to some power of $\sqrt{-g}$. It is found that
\beq
\label{eq:Gdet_power}
\left( \det \left[e^2 \sqrt{-g} 
(g^{\mu\al} g^{\nu\bt}- g^{\mu\bt} g^{\nu\al}) \right] \right)^{1/2} = 
e^{d(d-1)/2}
\left(
\sqrt{-g} \right)^{(d-1)(d-4)/4} \, .
\eeq
In four dimension this is a pure number and
is equal to $\left(e^2 \right)^3$.
Hence the dual theory functional looks exactly like the original 
functional times a constant. Classically the two actions are 
equivalent. At the quantum level the equivalence between the 
dual action given by eq. (\ref{eq:dualact}) and the original action 
$S_{EM}$ holds only on-shell \cite{FradkinD}. On doing perturbative
renormalization of the dual theory one gets,
\beq
\label{eq:beta_dual}
\frac{{\rm d}}{{\rm d}t} e^2 = a(\cdots) \, e^2 \, ,
\eeq
where $a$ has exactly the same parameter dependence as in 
Eq. (\ref{eq:e2beta}). 

Compatibility of the two equation (\ref{eq:e2beta} and \ref{eq:beta_dual})
implies that $a=0$. We have not explicitly introduced any gauge-fixing 
and the corresponding Faddev-Popov ghost functional. If one 
uses the same kind of gauge fixing conditions in the 
original and in the dual theory, then it is easy to show that the 
above argument goes through without any modification, again on-shell. 
Off-shell there will be gauge dependences and the quantum equivalence 
of the two theories will not be present, as also discussed in \cite{FradkinD}.
It should be noted that this argument doesn't depend on the regularization 
scheme or the gravity action, meaning both the $U(1)$ and its dual theory 
eq. (\ref{eq:e2beta} and \ref{eq:beta_dual}) are inferred in any regularization scheme.
But it is important to point out that this argument of self-duality being responsible 
for vanishing of the `$a$' term, is valid only on-shell.
But it is quite a compelling observation which forces one to 
hope that perhaps `$a$' term might be zero even off-shell. 

In the next section we explicitly show that at one-loop $a=0$
without any regularization scheme, independent of 
the gravity action and gauge fixing condition thereof.

\section{Arbitrary Gravity Action.}
\label{arbitGR}
%
By making use of the all spin projectors of a symmetric rank-2 tensor field,
the most general graviton propagator in a metric theory of gravity 
in an arbitrary harmonic type gauge fixing action about a 
flat background can be written as,
\beq
\label{eq:general_gravity_prop}
D^{\mn,\rho\sg}
= \frac{i}{\pd} \sum_i Y_i(p^2) P_i^{\mn,\rho\sg}
\eeq
where $i=\{2, 1, s, w, sw, ws\}$ and $Y_i(p^2)$ are 
propagators corresponding to various spin components. $P_2$ and $P_s$ 
have been given previously in eq. (\ref{eq:proj}) 
while the other spin projectors are: 
\bea
\label{eq:proj1}
&&
(P_1)_{\mn,\al\bt}
=\frac{1}{8}(T_{\mu\al} L_{\nu\bt}
+ T_{\mu\bt}L_{\nu\al} + T_{\nu\al}L_{\mu\bt}
+T_{\nu\bt}L_{\mu\al})\, ,
\notag \\
&&
(P_w)_{\mn,\al\bt}
=L_{\mn}L_{\al\bt} \, ,
\notag \\
&&
(P_{sw})_{\mn,\al\bt}
=\frac{1}{\sqrt{d-1}}T_{\mn}L_{\al\bt}\, ,
\notag \\
&&
(P_{ws})_{\mn,\al\bt}= \frac{1}{\sqrt{d-1}}L_{\mn}T_{\al\bt}.
\eea
Using this propagator of graviton we compute the one-loop 
diagram given in Fig. \ref{fig1}. The combined contribution of both the 
diagrams in arbitrary dimensions is given by, 
\bea
\label{eq:EA_grav}
&&
\G_{\rm Grav} = -\frac{i(d-4) A(d) }{8e^2} 
\int {\rm d}^dx \, {\rm tr} F^2 \, ,
\hspace{5mm}
{\rm where \,\,\,\,}
A(d) = \sum_i A_i
\int \frac{{\rm d}^dp}{\pd} 
Y_i(p^2) \, ,
\eea
and $A_2 =-(d-5)(d-2)(d+1)/4d(d-1)$, $A_1 =-(d-1)/8d$,
$A_s=(d^2 -12d+19)/4d(d-1)$, $A_{sw}=A_{ws} = \sqrt{d-1}/4d$,
and $A_w = -1/4d$.
We see explicitly that there is no gravitational 
contribution to the running of gauge coupling in four 
dimensions as the total contribution being proportional to $(d-4)$, vanishes in $d=4$.
It should be noted that this doesn't depend on the gravity
action, gauge parameters present in the theory and the 
regularization scheme, as we haven't performed the 
momentum integration and didn't have to specify the form and nature of
$Y_i(p^2)$. 
This further supports the fact observed in previous section,
namely, perhaps self-duality alone is responsible for the
vanishing of `$a$' term, while the result being independent of
the regularization scheme and the gravity action.
Lorentz covariance property of the background gravity 
metric $\eta_\mn$ was only used to arrive at eq. (\ref{eq:EA_grav}).
In eq. (\ref{eq:EA_grav}) we have just quoted the result, 
details of which will be presented elsewhere \cite{NarainA3}. 

A byproduct of the above computation is, that a subclass of higher-loop 
diagrams of tadpole and bubble type but with dressed graviton propagator 
also tend to cancel each other in four dimensions.

\section{Discussion and Conclusion.}
\label{discuss}

In this paper we study gauge fields coupled with higher-derivative 
gravity which has been shown to be perturbatively renormalizable 
\cite{Stelle, Fradkin, Moriya} and unitary \cite{NarainA1, NarainA2}.
The path-integral for such a system is completely well-defined
perturbatively in $(4-\ep)$ dimensional regularization scheme, 
which we use to show at one-loop there is no quantum gravity 
correction to beta function of gauge coupling. This theory has no 
quadratic divergences and is renormalizable to all loop, thereby 
evading the criticism raised in \cite{Donoghue4}.

Generally the beta function of the gauge coupling including the quantum 
gravity corrections consist of two kind of terms: `$a$' and `$b$' term as
shown in eq. (\ref{eq:beta_gauge_gr}). The `$a$' term is universal to all gauge 
theories and is independent of the matter content of the theories. Furthermore,
it only depends on the gravitational couplings, while the `$b$' term is not universal.
Here in this paper we studied just the nature of the `$a$' term to all loops in four dimensions.
At one loop it is found that the only quantum gravity contribution that enters 
the gauge beta function is `$a$' term, which is found to be zero for all 
gravity action with metric as the field variable. This is independent of the gauge 
group, gauge fixing condition and the regularization scheme. Being universal 
in nature and same for all gauge groups, we isolated it by studying just the abelian 
gauge theories coupled with gravity. Here using the duality transformation we studied 
the dual theory coupled with gravity, which is quantum mechanically equivalent 
to the original theory on-shell \cite{FradkinD}. In four dimensions the abelian theory is self dual, thus the 
beta function of the gauge coupling in the new theory has the same form as in the original theory
with the same coefficient `$a$', thereby implying the vanishing of `$a$' term. 
This is an all-loop argument which hold only on-shell and doesn't depend 
on the use of any particular gravity action. However for higher-derivative 
gravity we have a well defined path-integral. 

This has dramatic consequences. This means that 
photons interacting only with quantum fluctuations of metric,
propagate essentially as free particle at short distances namely no running 
of the charge coupling parameters. This is a consequence of self 
duality property of $U(1)$ gauge action. It should be mentioned 
that there is indeed a finite charge renormalization, however 
it is of no significance once we include other matter fields. For non-abelian 
gauge theories the same phenomenon manifests as there will not 
be any `$a$' term. However the running of the non-abelian gauge coupling is 
controlled by the `$b$' term.

In the absence of `$a$' term the only contribution to the beta function
is from `$b$' term. The leading one-loop contribution to the `$b$' term
in perturbation theory is completely independent of gravity parameters 
{\it i.e.} it is solely a consequence of charge interactions alone. At two-loops 
matter contributions alone do not alter asymptotic freedom 
of non-abelian gauge theories or Landau singularity for abelian gauge theories
\cite{Arason, Ford}.
Contributions which include gauge and gravity couplings exist at two loops 
and these can influence the beta functions for example in generating 
new fixed points or lines in the multi-dimensional gauge and gravity 
parameters space. This can open new avenues to our understanding of gravity 
and gauge field theories.

\paragraph*{Acknowledgements.}
%
GN would like to thank Romesh Kaul for useful discussions. 



\end{document}